\documentclass[%
 aip,
 pof,%
 amsmath,amssymb,
 reprint,%
 floatfix,
 tightenlines
]{revtex4-1}

\usepackage{graphicx}
\usepackage{dcolumn}
\usepackage{bm}
\usepackage{color}

\def\be { \begin{equation} }
\def\ee { \end{equation} }

\def\re {{ R_\lambda }}
\def\etal {{ \rm et al. }}

\def\uu { \mathbf{u} }

\def\uu {{ \mathbf{u} }}

\def\uB {{ \mathbf{B} }}

\definecolor{mygreen}{rgb}{0,0.5,0.}

\begin{document}
\title[Cancellation exponents in isotropic turbulence and MHD turbulence]
{Cancellation exponents in isotropic turbulence and magnetohydrodynamic turbulence}

\author{X.M. Zhai}%
\affiliation{
School of Aerospace Engineering, Georgia Institute of Technology, Atlanta, GA 30332, USA}%
\author{K.R. Sreenivasan}
\affiliation{Department of Mechanical and Aerospace Engineering, Department of Physics,
and Courant Institute of Mathematical Sciences,
New York University, New York, NY 10012, USA
}
\email{kr3@nyu.edu}
\author{P.K. Yeung}
\affiliation{
Schools of Aerospace Engineering and Mechanical Engineering, Georgia Institute of Technology, Atlanta, GA 30332, USA
}

\date{\today}

\begin{abstract}

Small scale characteristics of turbulence such as velocity gradients and vorticity fluctuate rapidly in magnitude and oscillate in sign. Much work exists on the characterization of magnitude variations, but far less on sign oscillations. While averages performed on large scales tend to zero because of the oscillatory character, those performed on increasingly smaller scales will vary with the averaging scale in some characteristic way. This characteristic variation at high Reynolds numbers is captured by the so-called cancellation exponent, which measures how local averages tend to cancel out as the averaging scale increases, in space or time. Past experimental work suggests that the exponents in turbulence depend on whether one considers quantities in full three-dimensional space or uses their one- or two-dimensional cuts. We compute cancellation exponents of vorticity and longitudinal as well as transverse velocity gradients in isotropic turbulence at Taylor-scale Reynolds number up to 1300 on $8192^3$ grids. The 2D cuts yield the same exponents as those for full 3D, while the 1D cuts yield smaller numbers, suggesting that the results in higher dimensions are more reliable. We make the case that the presence of vortical filaments in isotropic turbulence leads to this conclusion. This effect is particularly conspicuous in magnetohydrodynamic turbulence, where an increased degree of spatial coherence develops along the imposed magnetic field. 

%
\end{abstract}

\pacs{Valid PACS appear here}
\maketitle

\section{Introduction}

Small scale motions in fluid turbulence such as velocity gradients and vorticity exhibit fluctuations of positive and negative signs, both in space and time. If oscillations in sign continue to occur no matter how small a spatial or temporal interval is probed, a form of singularity can be said to exist. Even the smallest amount of averaging will cancel out the signal. This behavior is known as sign-singularity \citep{ott1992sign,du1994characterization,vainshtein1994scaling}. For all physical signals, this cancellation tendency occurs only over some range of averaging scales. 

Mathematically the idea is made clear with the introduction of a signed measure $\mu_i(l)$ at some scale $l$:
\be
\mu_i(l)=
\frac{\int_{Q_i(l)} d \mathbf{r} f(\mathbf{r}) }
{\int_{Q(L)} d \mathbf{r} |f(\mathbf{r})| }
\label{eq:signmeasure}
\ee
where $Q_i(l)$ denotes a hierarchy of disjoint subsets of size $l$ covering the entire domain $Q(L)$ of size $L$, and $f(\mathbf{r})$ is a scalar field with a zero mean value. The denominator is chosen to bound $\mu_i(l)$ between $[-1,1]$, thus making it a signed probability measure. The sum of the absolute values of all the signed probability measures gives rise to the partition function $\chi(l)$ defined as 
\be
\chi(l)=\sum_{Q_i(l)} |\mu_i(l)|.
\label{eq:partfn}
\ee
Since $\chi(l)=1$ if $f(\mathbf{r})$ is sign-definite, sign-singularity is readily reflected in non-unity $\chi(l)$, which is possible only when cancellations of opposite signs occur in the numerator of Eq.~\ref{eq:signmeasure}. Therefore, to measure the propensity of the quantity considered to cancel out when averaged over a region of space or an interval of time, the ``cancellation exponent" $\kappa$ is defined \citep{ott1992sign,du1994characterization} via
\be
\chi(l) \sim l^{-\kappa}
\label{eq:definition}
\ee
Clearly, sign-definite signals have $\kappa=0$.

\begin{figure}[!h]
\centering
\includegraphics{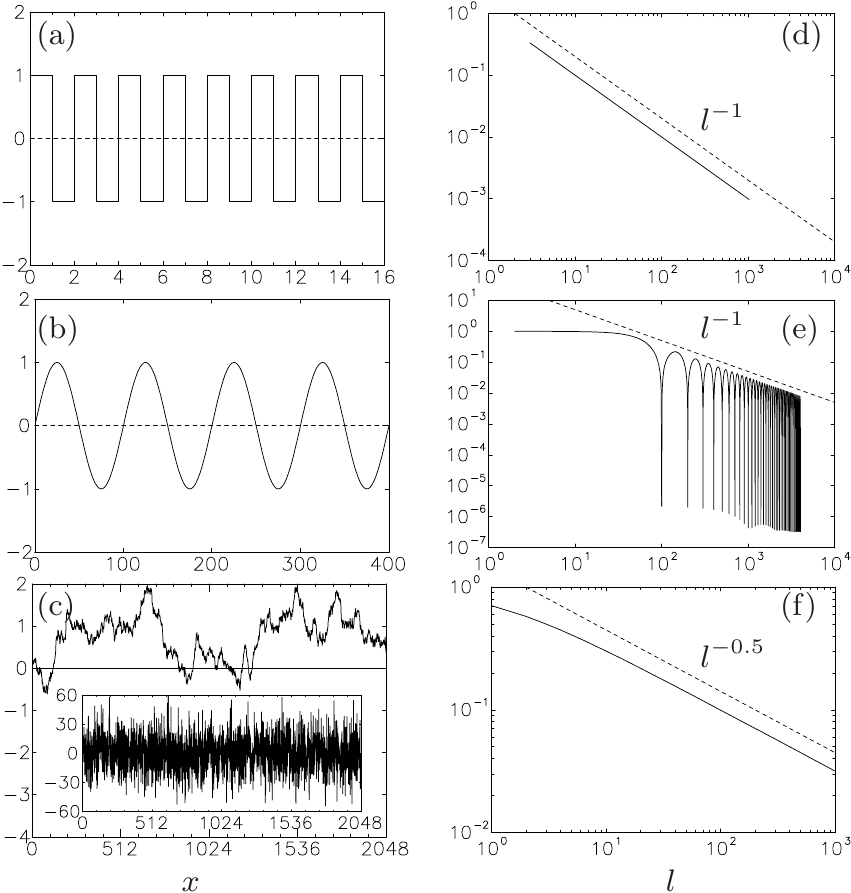}
\caption{
Signals of (a) a square wave with magnitude of unity; (b) a sinusoidal wave $y(x)=\sin\big( (2\pi/100) x \big)$ with a period of 100; (c) a standard Wiener process (Brownian motion) where the inset shows the ratio of step-wise increment and unit step size. For brevity, (a) (b) show only a few periods of the signal. Partition functions ($\chi(l)$) of signal (a-c) are shown in (d-f). Dashed lines mark power-law behaviors. 
}
\label{fig:demo}
\end{figure}

To help understand the properties of cancellation exponent, we show in Fig.~\ref{fig:demo} simple one-dimensional signals of square wave, sinusoidal wave and standard Wiener process (Brownian motion) as well as their partition functions. In Fig.~\ref{fig:demo} (a) for the square wave extending from $x=0$ to $2n$ (where $n$ is an integer), the signed measure $\mu_i(l)$ is zero when $l$ is even, and $\pm 1/(2n)$ when $l$ is odd. Since the number of disjoint subsets at size $l$ is $(2n)/l$, following Eq.~\ref{eq:partfn} the partition function $\chi(l)=\sum_{Q_i(l)} |\mu_i(l)| =(2n)/l \times 1/(2n)=1/l$ for odd values of $l$, and zero otherwise. As a result, plotting $\chi(l)$ as a function of odd numbers of $l$ only, Fig.~\ref{fig:demo} (d) shows that the cancellation exponent $\kappa=1$, which is known to be the case for non-differentiable signals \citep{vainshtein1994scaling}. In Fig.~\ref{fig:demo} (b) for the sinusoidal wave with a period of 100, the integral in the numerator of Eq.~\ref{eq:signmeasure} vanishes when the interval size $l$ takes multiples of the period and $\chi(l)$ is zero. Indeed, very small values of $\chi(l)$ are seen in Fig.~\ref{fig:demo} (e) for $l$ equal to any integral multiple of the period. Since finite numerical accuracy prevents the occurrence of exact zero, the small values of $\chi(l)$ appear as deep valleys. Furthermore, the signed measure $\mu_i(l)$ depends strongly on the interval size $l$, resulting in large variations of $\chi(l)$. The envelope, as expected, has a slope of $-1$. In Fig.~\ref{fig:demo} (c) for the standard Wiener process (Brownian motion), the ratio of the stepwise increment and step size (shown in the inset) is highly oscillatory, and is known \citep{bertozzi1994cancellation} to correspond to $\kappa=0.5$. Good match with $\kappa=0.5$ can be seen in Fig.~\ref{fig:demo} (f).

\begin{figure}[!h]
\centering
\includegraphics{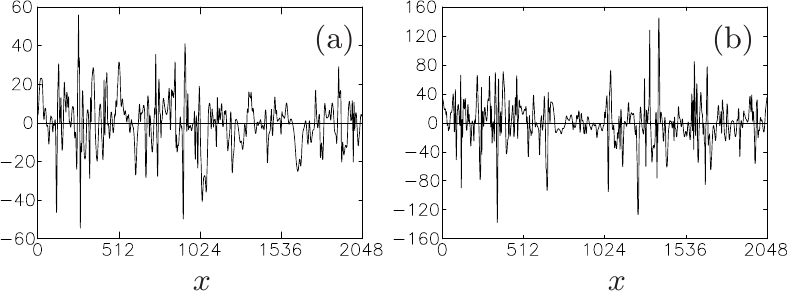}
\caption{
Line traces of (a) longitudinal velocity gradient $\partial u / \partial x$ and (b) vorticity component $\omega_z = \partial u/\partial y - \partial v/\partial x$ from a simulation of isotropic turbulence at $\re=400$ on a $2048^3$ grid.
}
\label{fig:linetrace}
\end{figure}

The examples constructed above show that even simple signals can be sign-singular. In fact, sign-singularity is ubiquitous in nature, such as in more sophisticated signals in magnetohydrodynamics (MHD) \citep{sorriso2002analysis, graham2005cancellation, martin2013cancellation}, solar activities \citep{carbone1997sign,consolini1999sign,carbone2010sign,sorriso2015sign}, geomagnetic field \citep{de1998sign}, helical flows \citep{imazio2010cancellation}, rotating turbulence \citep{horne2013sign} and aspects of classical turbulence \citep{ott1992sign,du1994characterization,vainshtein1994scaling}. As an example, Fig.~\ref{fig:linetrace} shows line traces of longitudinal velocity gradient $\partial u / \partial x$ and vorticity component $\omega_z = \partial u/\partial y - \partial v/\partial x$ from direct numerical simulations (DNS) of isotropic turbulence with a Taylor-scale Reynolds number $\re=400$. Both quantities oscillate strongly in sign, with vorticity exhibiting ostensibly greater intermittency than the longitudinal velocity gradient. We will discuss both signals in more detail in Sect. IV.

An \textit{unresolved} question in the study of cancellation exponents is whether and how different types of calculation methods affect the results. In particular, a natural question arises when sign cancellations are measured along lines (one-dimensional, 1D), in planar intersections (2D) or over three-dimensional (3D) volumes. While some theoretical results connect lower dimensional results with those in three-dimensions (see, e.g., Mandelbrot \citep{mandelbrot}, Sreenivasan \citep{sreeni.1991}, Vainshtein \etal \citep{vainshtein1994scaling}), it is not clear that they should work for real quantities in arbitrary flows. Experimental data analysis suggests that measurements over spatial extent of different dimensions are different. Past 2D measures of cancellation exponent for vorticity were larger, with $\kappa=0.85$\citep{vainshtein1994scaling}, than 1D measures, $\kappa=0.45$ \citep{ott1992sign} and $\kappa=0.6$ \citep{vainshtein1994sign}. These differences indicate that 1D measures are ``blind" to structures with dimensions less than two, and assessment in higher dimensions might be quite necessary for turbulent quantities \citep{vainshtein1994scaling}. However, partly due to difficulties of experimentally making measurements in 3D, a thorough comparison of cancellation exponents measured in all three dimensions has not been made. It follows that the underlying causes of the differences by measures of different dimensions have not been clearly identified.

One objective of this paper is to examine systematically how cancellation exponents measured in all three different dimensions differ. We use data from DNS of homogeneous isotropic turbulence, and compute cancellation exponents using 1D, 2D and 3D measures; indeed, quantifying sign oscillations in simulations can be more versatile than in experiments because of access in the latter to all the quantities of interest. In addition, the isotropic nature of turbulence in our simulations alleviates issues of large-scale anisotropy in experiments \citep{sreenivasan1995scaling}. Since the term ``cancellation exponent" was introduced about 25 years ago, great advances in computing power has now allowed us to examine the effects of Reynolds number as well, up to $\re=1300$ on $8192^3$ grids \citep{YZS.2015}. This is an important issue.

Three small-scale quantities---vorticity, longitudinal and transverse velocity gradients---are considered in this study. All three are found to have the same cancellation exponent of $2/3$ when measured in 2D and 3D, but 1D values are much smaller for transverse velocity gradients and vorticity, while they are close to $2/3$ for longitudinal velocity gradient even in 1D. For vorticity, there exists a relation between the cancellation exponent and the characteristic exponent for first order velocity increment \citep{vainshtein1994sign, vainshtein1994scaling,nikora2001intermittency}, and $\kappa$ is expected to be close to $2/3$. As a result, a cancellation exponent value of $2/3$ in 2D and 3D confirms dimensions higher than 1D are indeed necessary for quantifying sign cancellations, at least for vorticity. Our work suggests that the use of 1D measure in past experiments may be why $\kappa$ is underestimated, but it becomes necessary to understand this better. We provide an explanation from the perspective of the geometry of small-scale motions, and suggest the differences in cancellation exponents in different dimensions result from the prevalence of coherent structures in some preferred direction. The basic idea is that coherent structures are mostly composed of events of the same sign, and therefore 1D measures of sign oscillation, if taken along such structures, do not record effective sign cancellations and tend to give lower values of cancellation exponents. In comparison, the neighborhoods of coherent structures contain events of opposite signs, where sign cancellations occur in all directions. To demonstrate the idea, we consider MHD turbulence at low magnetic Reynolds numbers where the diffusion of the magnetic field is much stronger than the advective transport \cite{ZY.2018}. We observe that the substantial elongation of the vortical structures along the magnetic field is accompanied by a strong reduction of sign cancellations when 1D measure is used. This result suggests that a similar explanation holds also for vorticity in isotropic turbulence \citep{JWSR1993,tenchapters}.

The rest of the paper is organized as follows. In Sec.~II we outline the computational method and discuss measures in different dimensions. Our main results are presented in Sec.~III, where we show cancellation exponents measured in 1D, 2D and 3D for vorticity, longitudinal and transverse velocity gradients. In Sec.~IV, we show visualizations and cancellation exponents for low-$R_m$ MHD turbulence, and discuss the relationship between elongated structures and reduced values of cancellation exponents. Finally, in Sec.~V we present the conclusions and discuss the implications of the work.

\section{Computational method}



We perform DNS of the incompressible Navier-Stokes equations
\be
\partial \mathbf{u}/\partial t
+(\mathbf{u} \cdot \nabla) \mathbf{u}
= -\nabla (p/\rho) + \nu \nabla^2 \mathbf{u} + \mathbf{f}
\ee
where $\mathbf{u}$ is the solenoidal velocity field ($\nabla \cdot \mathbf{u}=0$), $p$ is pressure, $\rho$ is fluid density, $\nu$ is the kinematic viscosity and $\mathbf{f}$ is the forcing term that maintains a stationary state \citep{EP88, DY2010}. We use Fourier pseudo-spectral calculations \citep{rogallo} on a periodic domain of size $(2\pi)^3$ with an explicit second order Runge-Kutta integration in time. A combination of phase-shifting and truncation is used to reduce aliasing errors, where the highest resolved wavenumber $k_{max}=\sqrt{2}N/3$ and $N$ is the number of grid points in one dimension. Typical spatial resolution, expressed by $k_{max}\eta$, is around $1.5$ for simulations aimed at higher Reynolds numbers \citep{IGK2009}. Recently Yeung \etal \citep{YSP.2018} pointed out that at higher Reynolds number, more stringent spatial and temporal resolution are necessary. For the data analysis in this paper, we have used datasets with improved resolution of $k_{max\eta} \geq 2$ over a wide range of Taylor-scale Reynolds number $\re=140$ to 1300, as summarized in Table~\ref{tab:simulation}.

\begin{table}[h]
\begin{center}
\begin{tabular}{c c c c c c }
\hline
$R_\lambda$ & 140 & 240 & 400 & 650 & 1300 \\
$k_{max}\eta$ & 5.6 & 5.6 & 2.7 & 2.7 & 2 \\
$N$ & 1024 & 2048 & 2048 & 4096 & 8192 \\
$N_R$ & 8 & 14 & 16 & 12 & 6 \\
\end{tabular}
\end{center}
\caption{Data sets of isotropic turbulence used in the analysis. $R_\lambda$ is the Taylor-scale Reynolds number. Spatial resolution is denoted by $k_{max}\eta$. $N$ is the number of grid points along each side of the cubic domain. $N_R$ denotes the number of realizations used for ensemble-averaging.}
\label{tab:simulation}
\end{table}

For MHD turbulence, motions of electrically-conducting fluids under an external magnetic field $\mathbf{B_0}$ produce a current, which induces a secondary fluctuating magnetic field $\mathbf{b}$, and also gives rise to the Lorentz force that modifies the momentum equation. At low magnetic Reynolds number ($R_m$), the induced fluctuating magnetic field is quickly diffused away by strong magnetic diffusion and is therefore much weaker (i.e. $|\mathbf{b}|\ll|\mathbf{B_0}$).
Moreover with the quasi-static approximations at $R_m \ll 1$, we only need to consider how the velocity field is affected by the magnetic field. Specifically the momentum equation becomes
\begin{align}
\partial\uu/\partial t
+ (\uu\cdot\nabla) \uu 
= &- ({1}/{\rho}) \nabla (p+ {B_0^2}/{2\mu}) + \nu\nabla^2\uu
\nonumber \\
&- ({\sigma}/{\rho})
[(\uB_0\cdot\nabla)^2(\nabla^{-2}\uu)] \
\label{eq:lowrm}
\end{align}
which can be readily transformed to Fourier space. Numerically the Lorentz term (the last term in Eq.~\ref{eq:lowrm}) is treated exactly via an integrating factor. Unlike in isotropic turbulence, forcing is not applied in low-$R_m$ MHD turbulence simulations to avoid interference with the physics of the Lorentz force, which acts at all scales. The turbulence field is initialized with a model energy spectrum, and is then allowed to take on Navier-Stokes dynamics during its decay. The magnetic field is activated when the non-Gaussian feature of the velocity field is well developed. More details of the simulations can be found in Zhai \& Yeung \citep{ZY.2018}. 

One key element of the analysis is to contrast cancellation exponent $\kappa$ obtained from 1D, 2D and 3D measures. As a result the meaning of $Q_i(l)$ and $Q(L)$ in Eq.~\ref{eq:signmeasure} depends on the dimensionality of the measure: $Q_i(l)$ can come from line segments (1D), square areas (2D) and cubes (3D), all with edge length of $l$; $Q(L)$ can come from box length $L$, side area $L^2$ and volume $L^3$. The use of 2D domain decomposition \citep{DYP2008} in the simulations poses computational challenges for 2D and 3D measures as data needed for evaluating Eq.~\ref{eq:signmeasure} may be distributed among multiple processors, but strategies such as prefix sums \citep{iyer.thesis} have been adopted to reduce computation and communication loads. To allow for direct comparisons with experiments \citep{vainshtein1994scaling}, 2D measures are recovered through the application of Stokes theorem. Taking vorticity component as an example, the circulation $\Gamma_A(l)$ of the velocity field $\mathbf{v}$ around a closed loop $\mathbf{s}$ surrounding an area $A=l^2$ is
\be
\Gamma_A(l)=\oint \mathbf{v} d \mathbf{s} =
\int_A \mathbf{\omega} \cdot \mathbf{n} d A
\label{eq:circulation}
\ee
If the circulation scales as $\langle |\Gamma_A(l)|^q \rangle \sim l^{\alpha_q}$ (where $q$ is any real number), it is shown in Ref. 
[3]
that $\alpha_q = (2-\kappa)q-(D-D_q)(q-1)$, where the space dimension $D=3$ and $D_q$ is the generalized dimension \citep{mandelbrot,hentschel1983}. For $q=1$, clearly $\kappa=2-\alpha_1$.

\section{Cancellation exponents in homogeneous isotropic turbulence}

\begin{figure}[!h]
\centering
\includegraphics{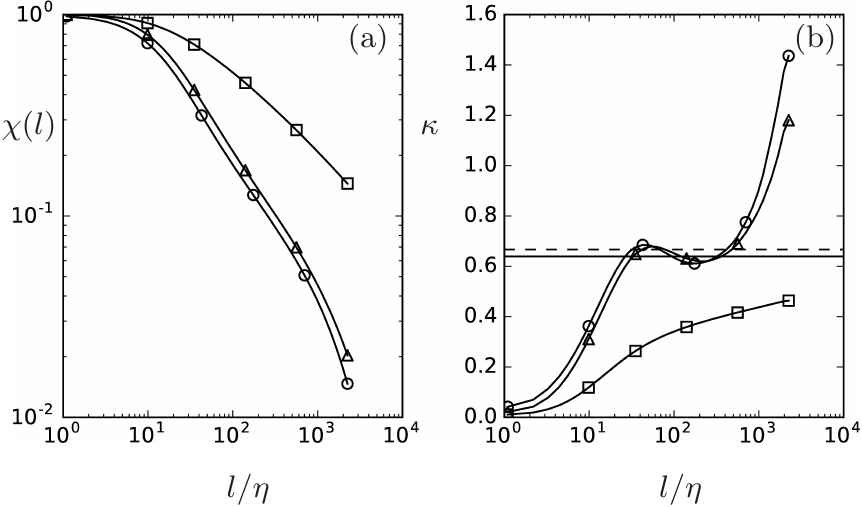}
\caption{(a) Partition function and (b) cancellation exponent $\kappa$ (see Eqs.~\ref{eq:partfn} and 3) for measures of $1D$ ($\square$), $2D$ ($\triangle$) and $3D$ ($\bigcirc$). Horizontal dashed line marks $2/3$, and solid line marks 0.639 as a result of log-normal correction for intermittency with the exponent $\mu=0.25$. Data are ensemble averaged at $R_\lambda=650$, $4096^3$.
}
\label{fig:omega}
\end{figure}

Since cancellation exponents are simply the scaling exponents of the partition function, it is instructive to plot both quantities side by side, as shown in Fig.~\ref{fig:omega} for vorticity measured in 1D and 2D cuts as well in 3D in homogeneous isotropic turbulence at $\re=650$. In the spirit of the inertial range, the scaling of the partition function is sought in a certain range of scales. Instead of fitting straight lines in the log-log plots of $\chi(l)$, the plateau regions in the local slopes $-d\, log\, [\chi(l)] / d\, log\, (l)$ are used to obtain the value of cancellation exponent. For small values of $l/\eta$ viscosity smooths the signals and weakens sign cancellations rendering $\chi(l)$ close to 1, as confirmed in Fig.~\ref{fig:omega} (a). Figure~\ref{fig:omega} (b) shows that plateaus indeed exist for 2D and 3D measures at around $50 < l/\eta < 400$, which is consistent with the inertial range identified in previous work \citep{ISY.2017}. Furthermore 2D and 3D measures give similar values of $\kappa \approx 2/3$, larger than what one may infer from the 1D measure, which does not show a convincing scaling in the first place.

The relationship between cancellation exponent and other scaling exponents in turbulence \citep{vainshtein1994sign,vainshtein1994scaling} can be used to explain the value of $2/3$. Following Vainshtein et al., \citep{vainshtein1994scaling}, we consider the generalized structure function at order $q$ where $q$ is any real number. In the inertial range, $\langle |\Delta u|^q \rangle \sim l^{\zeta_q}$ and the scaling exponent $\zeta_q$ is related to the cancellation exponent $\kappa$ by
\be
\zeta_q=(1-\kappa)q-(D-D_q)(q-1)
\label{eq:zeta}
\ee
where dimension of space $D=3$ and $D_q$ is the generalized dimension \citep{mandelbrot,hentschel1983}. For $q=1$, we have
\be
\zeta_1=1-\kappa
\label{eq:zeta1}
\ee
If the effects of intermittency were neglected,
Kolmogorov's hypothesis \citep{K41} gives $\zeta_1=1/3$ and thus $\kappa=2/3$; whereas refined similarity hypothesis \citep{K62} gives $\zeta_1=0.361$ and $\kappa=0.639$ (using lognormal correction with intermittency exponent $\mu=0.25$ \citep{SK.1993}). We use lognormality as an example of intermittency models without necessarily endorsing it. It is clear from Fig.~\ref{fig:omega} (b) that cancellation exponents $\kappa \approx 2/3$ measured in 2D and 3D are in good agreement with the relations above, but not for the 1D measure. In short, our data suggests that cancellation exponents obtained from 2D and 3D measures are consistent with theoretical expectations.

\begin{figure}[!h]
\centering
\includegraphics{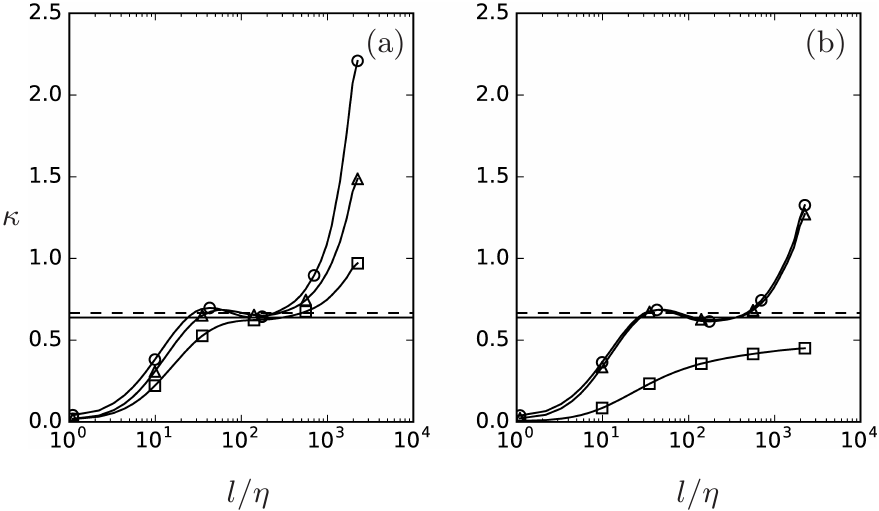}
\caption{Cancellation exponent $\kappa$ of (a) longitudinal velocity gradients and (b) transverse velocity gradients for measures of $1D$ ($\square$), $2D$ ($\triangle$) and $3D$ ($\bigcirc$). Horizontal dashed and solid lines mark $2/3$ and 0.639, respectively, as in Fig.~\ref{fig:omega}. Data are ensemble averaged at $R_\lambda=650$, $4096^3$.
}
\label{fig:vg}
\end{figure}

To see whether measurements of different dimensions have an effect on other small scale quantities, we perform similar calculations for longitudinal and transverse velocity gradients and show the cancellation exponents in Fig.~\ref{fig:vg}. While transverse velocity gradients behave similarly to vorticity, longitudinal velocity gradients are seen to have the same cancellation exponents $\kappa \approx 2/3$ for all three dimensions. To interpret the value of $2/3$ of the longitudinal velocity gradient using 1D measure, we note \citep{bertozzi1994cancellation} that the H{\"o}lder exponent $\alpha$ (for the first order structure function) of the velocity increment is related to the cancellation exponent of the velocity derivative $\kappa_1$ as $\kappa_1 = 1 - \alpha$. Again by the Kolmogorov hypothesis \citep{K41}, the H{\"o}lder exponent $\alpha=1/3$ and $\kappa_1 = 2/3$.  Fig.~\ref{fig:vg} (a) suggests that this relation holds in 2D and 3D. The close similarity of cancellation exponents in transverse velocity gradients (Fig.~\ref{fig:vg} (b)) and in vorticity (Fig.~\ref{fig:omega}) is perhaps not surprising, as vorticity is composed of algebraic combinations of transverse velocity gradients.

\begin{table}[h]
\begin{center}
\begin{tabular}{l l l}
D & $\kappa$ & experimental method \\
\hline
1 & 0.45 &
\begin{tabular}{@{}l@{}}1D cuts of one vorticity component \\
behind cylinder wake \citep{ott1992sign} \end{tabular} \\
1 & 0.6 &
\begin{tabular}{@{}l@{}}velocity difference over variable time interval \\ 
$\Delta u/\Delta t$ in atmospheric flow \citep{ott1992sign, vainshtein1994sign} 
\end{tabular} \\
2 & 0.85 &
2D circulation data behind cylinder wake 
\citep{vainshtein1994scaling,sreenivasan1995scaling} \\
\end{tabular}
\end{center}
\caption{
Cancellation exponent $\kappa$ for vorticity obtained from past experiments,
using 1D ($D=1$) and 2D ($D=2$) measurements.
}
\label{tab:experiment}
\end{table}

It is helpful now to comment on the past data. Table~\ref{tab:experiment} lists the cancellation exponents $\kappa$ of vorticity measured in past experiments with a brief summary of the experimental method. The lower value of $\kappa=0.45$ is likely due to the use of 1D measure, as reproduced in Fig.~\ref{fig:omega}. The data for $\kappa=0.6$ comes from atmospheric flow measurements where velocity differences over variable sampling time interval (i.e. $\Delta u/\Delta t$) were actually measured \citep{ott1992sign}. Yet, the data were interpreted as vorticity statistics by Vainshtein \etal \citep{vainshtein1994sign}, who considered the one-dimensional case and invoked Taylor's hypothesis. Strictly speaking, the $\kappa=0.6$ result is a confirmation of the relation between the H{\"o}lder exponent of a signal and its derivative \citep{bertozzi1994cancellation}, similar to results of $\kappa \approx 2/3$ in longitudinal velocity gradients in Fig.~\ref{fig:vg} (a), rather than vorticity. The $\kappa=0.85$ result measured from 2D circulation data behind cylinder wake \citep{sreenivasan1995scaling} is qualitatively consistent with a larger cancellation exponents by 2D and 3D measures from our numerical simulations (Fig.~\ref{fig:omega}), but our numerical simulations do not have the anisotropy of the cylinder wake.

\begin{figure*}
\centering
\includegraphics{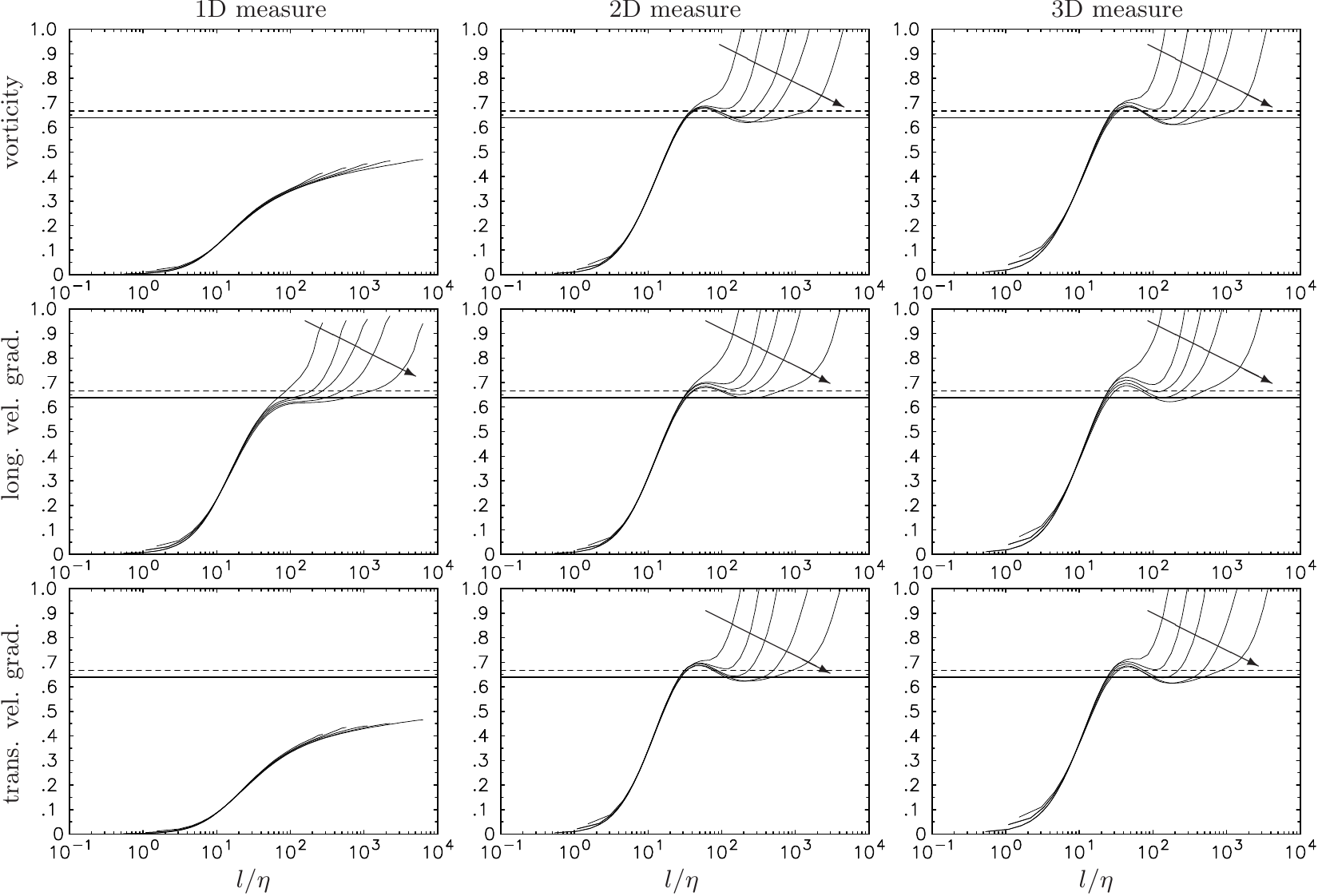}
\caption{Reynolds number dependence of cancellation exponents for vorticity (top row), longitudinal velocity gradients (middle row), and transverse velocity gradients (bottom row). From left to right, different columns denote measures in 1D, 2D and 3D; with  Reynolds number $\re$ increasing in the direction of the arrow for 140, 240, 400, 650 and 1300. Horizontal dashed and solid lines mark $2/3$ and 0.639, respectively, as in Fig.~\ref{fig:omega}. 
}
\label{fig:re}
\end{figure*}

We also study the Reynolds number dependence of cancellation exponents. Figure \ref{fig:re} shows cancellation exponents computed for vorticity, longitudinal and transverse velocity gradients using 1D, 2D and 3D measures, from $\re=140$ to $1300$. More extensive scaling range appears at the higher Reynolds number, as one should expect, and coincides with the inertial range reported previously \citep{ISY.2017}. The general observation is that 2D and 3D cancellation exponents for vorticity and transverse velocity gradients give similar values but larger than 1D measure, which does not show convincing plateaus. In comparison for longitudinal velocity gradients, measures of different dimensions give similar cancellation exponents. We note that the plateau is not perfect, but the values oscillate around $\kappa \approx 2/3$, perhaps due to a conspicuous bottleneck effect \cite{DS2010b}. 

\section{Cancellation exponent in low-$R_m$ MHD turbulence}

The results so far suggest that 1D measures of cancellation exponent of vorticity take smaller values than those obtained from 2D and 3D measures, and the question is why. Vainshtein \etal \cite{vainshtein1994scaling} argued that 1D measure is ``blind" to certain type of geometric structures. Martin \etal \cite{martin2013cancellation} also argued that coherent turbulence structures are ``smooth regions embedded in a highly fluctuating field" and as a result ``their presence and characteristics will influence the statistical properties of the scale-dependent changes of the sign". Both arguments suggest a connection between cancellation exponents and the structures of turbulent motions.

Consider a turbulent structure of any sign-oscillating quantity that has considerable coherency in one dimension (say, the $x$-direction). Such a structure can be a 1D filament or a flat sheet that extends in the $x$-direction. It is expected that signals of the same sign are embedded in the coherent structure, whereas signals of opposite signs can be found in the neighborhood of the structure (if signals of the \textit{same} sign are found in the neighborhood as well, the increased degree of coherency would extend beyond one dimension). When a 1D measure is used to quantify sign oscillations along the coherent structure, the persistence of the same sign reduces sign cancellations, leading to a smaller cancellation exponent. In contrast, 2D and 3D measures have more room in other dimension(s) for cancellation to take place, thus resulting in larger cancellation exponents.

Following this reasoning, the prevalence of vortex filaments in high Reynolds number isotropic turbulence may be thought to lead to differences in cancellation exponents measured in 1D versus higher dimensions. However, sign oscillation measures are taken along the coordinate axis while vortex filaments are randomly oriented in space. As a result there is only a fraction of coherent filaments that align with the grid axis in any realization and affect the cancellation exponent in the way described above. It is not clear if the use of many more realizations will solve this problem, but we can clarify if our reasoning is right. To this end, we consider low-$R_m$ MHD turbulence (where $R_m$ is the magnetic Reynolds number) in which the vortex structures are forced to be along a chosen coordinate axis---since vortical structures are known to grow preferentially along the magnetic field direction \citep{zikanov.1998, reddy.2014b}.

As noted earlier in Sec.~II,
if we assume $R_m\ll 1$, the induced secondary magnetic field is much weaker than the uniform mean magnetic field $\uB_0$, and we only need to focus on how the velocity field is affected by the magnetic field. As the magnetic field is applied to isotropic turbulence, integral length scales grow strongly along the magnetic field direction while the small scales of turbulence depart from local isotropy. Specifically, the velocity gradients are weakened in the direction of the magnetic field while the vorticity component becomes stronger and elongated. Zhai \& Yeung \citep{ZY.2018} have shown that an elongated domain is critical for alleviating confinement effects that arise from the use of periodic domains. Yet to focus on how elongated vortical structures affect the cancellation exponents measured in different dimensions, we use results on cubic domains of size $(2 \pi)^3$ on $2048^3$ grids.

\begin{figure*}
\centering
\includegraphics{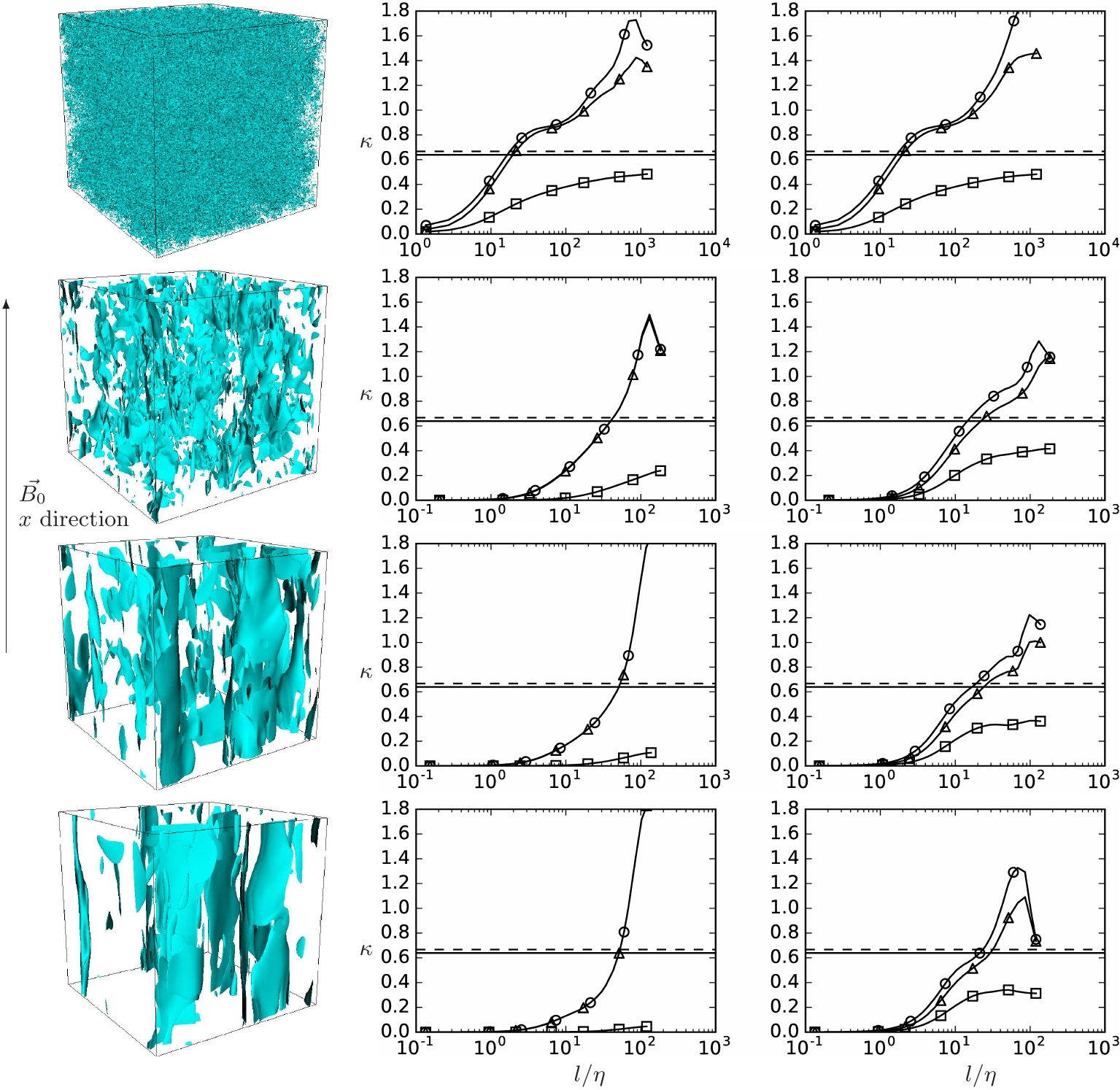}
\caption{Left column: visualization of normalized enstrophy $\Omega / \langle \Omega \rangle=5$ in MHD turbulence with the magnetic field along the $x$-direction (vertical); middle column: cancellation exponent of the $x$-component vorticity $\omega_x$; right column: averaged cancellation exponent of $\omega_y$ and $\omega_z$. Measures used are $1D$ ($\square$), $2D$ ($\triangle$) and $3D$ ($\bigcirc$). From top to bottom, $t/(\mathcal{L}/\mathcal{U})=0$, 12, 24 and 36.
}
\label{fig:mhd}
\end{figure*}

Figure~\ref{fig:mhd} shows the evolution of normalized enstrophy $\Omega=|\mathbf{\omega}|^2$ as well as the cancellation exponents of $\omega_x$ and the average of exponents of $\omega_y$ and $\omega_z$. The time is normalized by the ratio between integral length scale $\mathcal{L}$ and root-mean-square velocity $\mathcal{U}$, both computed at the instant of the  application of the magnetic field (top row). At $t/(\mathcal{L}/\mathcal{U})=0$, small vortex filaments are space-filling, and similar values of cancellation exponents for $\omega_x$ and averaged $\omega_y$ and $\omega_z$ confirm that isotropy holds to an acceptable level. At this low $\re=98$, cancellation exponents are qualitatively similar to those observed at $\re=140$ for forced isotropic turbulence (compare the first row for vorticity in Fig.~\ref{fig:re}). As turbulence decays, vortical structures become increasingly elongated along the magnetic field direction ($x$-direction). 
Moreover as the flow evolves the range of scales (measured by the ratio $l/\eta$) decreases because $\eta$ increases in time. The most notable change is that the 1D result of cancellation exponent for $\omega_x$ becomes significantly smaller than those for 2D and 3D measures (middle column). Yet the lack of any plateau in $\kappa$ suggests that $\omega_x$ is not sign-singular. For completeness, we note in the right column that for $\omega_y$ and $\omega_z$, a clear plateau is only seen for 1D measure at intermediate and large scales. The inflections of the curves by 2D and 3D measures mimic those in Fig.~\ref{fig:re}, but better-defined plateaus may form at higher Reynolds numbers.

\begin{figure}
\centering
\includegraphics{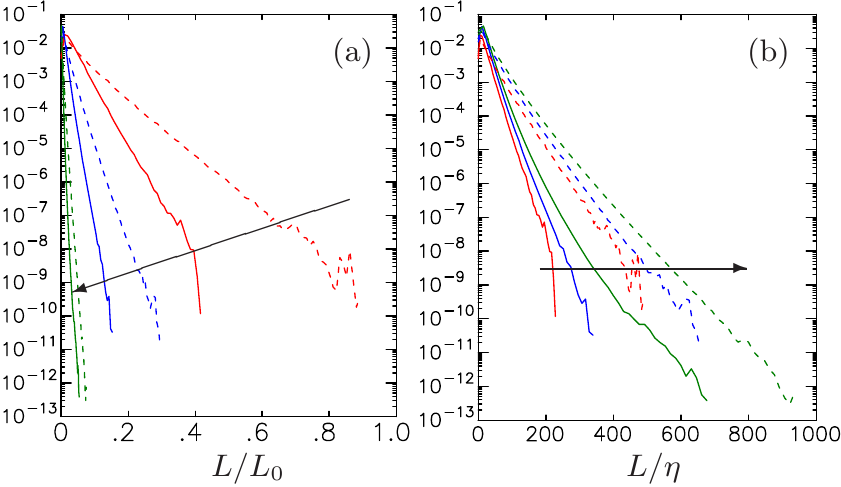}
\caption{
PDFs of the interval length of two signals over which the signal retains the same sign in one dimension. Solid curves denote longitudinal velocity gradients and dashed curves denote transverse velocity gradients. Arrows point along increasing $\re=140$ (red), 400 (blue) and 1300 (green). In (a) $L$ is normalized by the domain length ($L_0$) while in (b) it is normalized by the Kolmogorov length scale ($\eta$).
}
\label{fig:cohlenpdf}
\end{figure}

The example of low-$R_m$ MHD turbulence confirms that increased degree of coherence in turbulent structures can effectively reduce sign cancellations when the 1D measure along a specific direction is used. The coherent structures in low-$R_m$ MHD turbulence are sheets elongated preferentially along one direction, whereas they are filaments for vorticity in isotropic turbulence. To examine the degree of coherency in one-dimension for longitudinal and transverse velocity gradients, we show in Fig.~\ref{fig:cohlenpdf} the probability density functions (PDF) of the interval length $(L)$ over which two one-dimensional signals retain their sign in one direction. The PDF of $L/L_0$ (the interval length normalized by the domain size $L_0$) in Fig.~\ref{fig:cohlenpdf} (a) shows that, at higher Reynolds number, it is less likely for both longitudinal and transverse velocity gradients to maintain the same sign over extended scales, which is in agreement with the notion that turbulence tends to rupture coherent structures. When normalized on the length $L/\eta$, Fig.~\ref{fig:cohlenpdf} (b) shows that it is more likely for transverse velocity gradients to form longer coherent structures than longitudinal gradients. Therefore, as longitudinal velocity gradients are more fragmented, the less coherent structures in 1D longitudinal velocity gradients have minimal effects on cancellation exponents measured in 1D, 2D and 3D.

\section{Discussion and conclusions}

We have revisited the concept of cancellation exponents using high-resolution DNS of homogeneous isotropic turbulence up to $\re=1300$ on $8192^3$ grids. A highlight of this work is the computation of cancellation exponents in 1D and  2D cuts and their comparisons with those of the full 3D quantities. The 3D measures are hardly attainable in experiments. This work has allowed us to resolve conflicts in previous data and finally allow a direct assessment on whether measures of dimensions higher than unity are needed to measure cancellation exponents for turbulence processes \citep{vainshtein1994scaling}. Specifically, our results show that the answer depends on the quantity in question. For vorticity and transverse velocity gradients, 2D and 3D measures of cancellation exponents are close to $\kappa \approx 2/3$, and larger than the 1D measure. However, longitudinal velocity gradients have similar cancellation exponents of $\kappa \approx 2/3$ regardless of the dimensionality of the measure. By invoking connections to exponents of generalized structure functions\citep{vainshtein1994scaling}, we show that for vorticity the cancellation exponent should indeed be close to $2/3$; this reveals that the 1D measure is not sufficient. Results from simulations provide more insight on past experimental work. Specifically, in past experiments that quantify sign-oscillations in vorticity, the lower value of $\kappa=0.45$ is likely due to the fact that 1D measure was used; on the other hand, a value of $\kappa=0.6$, close to $2/3$, obtained for longitudinal velocity gradients suggests that it is not as sensitive to the dimensionality. The suspected reason for the discrepancies for quantities in different directions, as well as 1D, 2D and 3D measures, is the existence of persistent coherent structures.

To better understand the reasons underlying these differences, we have analyzed cancellation exponents of vorticity in low-$R_m$ MHD turbulence. Compared to forced isotropic turbulence where vortex filaments are randomly oriented in space, in low-$R_m$ MHD turbulence vortical structures grow preferentially along the magnetic field direction. As a result, better alignment of elongated coherent structures with the direction of 1D measure allows us to assess whether increased degree of coherency leads to weakened sign-cancellation. Quantitatively, 1D measures of cancellation exponents are substantially reduced as elongated coherent vortical structures grow in the form of 1D filaments or 2D sheets, as confirmed by qualitative visualizations. It is thus very plausible that in homogeneous isotropic turbulence elongated vortical structures in the form of filaments are responsible for smaller cancellation exponents measured in 1D. In comparison, structures of longitudinal velocity gradients are more fragmented, leading to similar cancellation exponents regardless of the dimensionality of the measure.

We briefly discuss two implications of this work. First, our results suggest that 1D measures can give misleading results for certain oscillatory quantities in 3D. In comparison, measurements in 2D and 3D yield more robust results that are less biased by the presence of structures with an increased degree of spatial coherence. As a result, interpretation of experimental results obtained using 1D measures requires extra caution \citep{horne2013sign}, and higher order measures should be preferred as a rule. Second, the demonstrable correlation between the structures and cancellation exponents allows the use of cancellation exponent as a convenient tool to monitor geometrical changes. For example, we may expect that the variations in cancellation exponents may be useful to monitor when geometrical changes in magnetospheric substorms \citep{consolini1999sign} and solar flares \citep{sorriso2015sign} are of interest.

\section*{Acknowledgments}

This work is supported by Grants ACI-1036170 and 1640771 under the Petascale Resource Allocations Program, and Grant CBET-1510749 under the Fluid Dynamics Program, both funded by the National Science Foundation (NSF). The computations and data analyses reported in this paper were performed using BlueWaters at the National Center for Supercomputing Applications (NCSA), University of Illinois at Urbana-Champaign, and the Texas Advanced Computation Center (TACC) of the University of Texas at Austin under the XSEDE program supported by NSF. The authors acknowledge helpful discussions with Kartik P. Iyer.



\begin{thebibliography}{38}%
\makeatletter
\providecommand \@ifxundefined [1]{%
 \@ifx{#1\undefined}
}%
\providecommand \@ifnum [1]{%
 \ifnum #1\expandafter \@firstoftwo
 \else \expandafter \@secondoftwo
 \fi
}%
\providecommand \@ifx [1]{%
 \ifx #1\expandafter \@firstoftwo
 \else \expandafter \@secondoftwo
 \fi
}%
\providecommand \natexlab [1]{#1}%
\providecommand \enquote  [1]{``#1''}%
\providecommand \bibnamefont  [1]{#1}%
\providecommand \bibfnamefont [1]{#1}%
\providecommand \citenamefont [1]{#1}%
\providecommand \href@noop [0]{\@secondoftwo}%
\providecommand \href [0]{\begingroup \@sanitize@url \@href}%
\providecommand \@href[1]{\@@startlink{#1}\@@href}%
\providecommand \@@href[1]{\endgroup#1\@@endlink}%
\providecommand \@sanitize@url [0]{\catcode `\\12\catcode `\$12\catcode
  `\&12\catcode `\#12\catcode `\^12\catcode `\_12\catcode `\%12\relax}%
\providecommand \@@startlink[1]{}%
\providecommand \@@endlink[0]{}%
\providecommand \url  [0]{\begingroup\@sanitize@url \@url }%
\providecommand \@url [1]{\endgroup\@href {#1}{\urlprefix }}%
\providecommand \urlprefix  [0]{URL }%
\providecommand \Eprint [0]{\href }%
\providecommand \doibase [0]{http://dx.doi.org/}%
\providecommand \selectlanguage [0]{\@gobble}%
\providecommand \bibinfo  [0]{\@secondoftwo}%
\providecommand \bibfield  [0]{\@secondoftwo}%
\providecommand \translation [1]{[#1]}%
\providecommand \BibitemOpen [0]{}%
\providecommand \bibitemStop [0]{}%
\providecommand \bibitemNoStop [0]{.\EOS\space}%
\providecommand \EOS [0]{\spacefactor3000\relax}%
\providecommand \BibitemShut  [1]{\csname bibitem#1\endcsname}%
\let\auto@bib@innerbib\@empty
\bibitem [{\citenamefont {Ott}\ \emph {et~al.}(1992)\citenamefont {Ott},
  \citenamefont {Du}, \citenamefont {Sreenivasan}, \citenamefont {Juneja},\
  and\ \citenamefont {Suri}}]{ott1992sign}%
  \BibitemOpen
  \bibfield  {author} {\bibinfo {author} {\bibfnamefont {E.}~\bibnamefont
  {Ott}}, \bibinfo {author} {\bibfnamefont {Y.}~\bibnamefont {Du}}, \bibinfo
  {author} {\bibfnamefont {K.~R.}\ \bibnamefont {Sreenivasan}}, \bibinfo
  {author} {\bibfnamefont {A.}~\bibnamefont {Juneja}}, \ and\ \bibinfo {author}
  {\bibfnamefont {A.~K.}\ \bibnamefont {Suri}},\ }\bibfield  {title} {\enquote
  {\bibinfo {title} {{Sign-singular measures: Fast magnetic dynamos, and
  high-Reynolds-number fluid turbulence}},}\ }\href@noop {} {\bibfield
  {journal} {\bibinfo  {journal} {Phys. Rev. Lett.}\ }\textbf {\bibinfo
  {volume} {69}},\ \bibinfo {pages} {2654} (\bibinfo {year}
  {1992})}\BibitemShut {NoStop}%
\bibitem [{\citenamefont {Du}, \citenamefont {T{\'e}l},\ and\ \citenamefont
  {Ott}(1994)}]{du1994characterization}%
  \BibitemOpen
  \bibfield  {author} {\bibinfo {author} {\bibfnamefont {Y.}~\bibnamefont
  {Du}}, \bibinfo {author} {\bibfnamefont {T.}~\bibnamefont {T{\'e}l}}, \ and\
  \bibinfo {author} {\bibfnamefont {E.}~\bibnamefont {Ott}},\ }\bibfield
  {title} {\enquote {\bibinfo {title} {Characterization of sign singular
  measures},}\ }\href@noop {} {\bibfield  {journal} {\bibinfo  {journal}
  {Physica D}\ }\textbf {\bibinfo {volume} {76}},\ \bibinfo {pages} {168--180}
  (\bibinfo {year} {1994})}\BibitemShut {NoStop}%
\bibitem [{\citenamefont {Vainshtein}\ \emph {et~al.}(1994)\citenamefont
  {Vainshtein}, \citenamefont {Sreenivasan}, \citenamefont {Pierrehumbert},
  \citenamefont {Kashyap},\ and\ \citenamefont
  {Juneja}}]{vainshtein1994scaling}%
  \BibitemOpen
  \bibfield  {author} {\bibinfo {author} {\bibfnamefont {S.~I.}\ \bibnamefont
  {Vainshtein}}, \bibinfo {author} {\bibfnamefont {K.~R.}\ \bibnamefont
  {Sreenivasan}}, \bibinfo {author} {\bibfnamefont {R.~T.}\ \bibnamefont
  {Pierrehumbert}}, \bibinfo {author} {\bibfnamefont {V.}~\bibnamefont
  {Kashyap}}, \ and\ \bibinfo {author} {\bibfnamefont {A.}~\bibnamefont
  {Juneja}},\ }\bibfield  {title} {\enquote {\bibinfo {title} {Scaling
  exponents for turbulence and other random processes and their relationships
  with multifractal structure},}\ }\href@noop {} {\bibfield  {journal}
  {\bibinfo  {journal} {Phys. Rev. E}\ }\textbf {\bibinfo {volume} {50}},\
  \bibinfo {pages} {1823} (\bibinfo {year} {1994})}\BibitemShut {NoStop}%
\bibitem [{\citenamefont {Bertozzi}\ and\ \citenamefont
  {Chhabra}(1994)}]{bertozzi1994cancellation}%
  \BibitemOpen
  \bibfield  {author} {\bibinfo {author} {\bibfnamefont {A.~L.}\ \bibnamefont
  {Bertozzi}}\ and\ \bibinfo {author} {\bibfnamefont {A.~B.}\ \bibnamefont
  {Chhabra}},\ }\bibfield  {title} {\enquote {\bibinfo {title} {Cancellation
  exponents and fractal scaling},}\ }\href@noop {} {\bibfield  {journal}
  {\bibinfo  {journal} {Phys. Rev. E}\ }\textbf {\bibinfo {volume} {49}},\
  \bibinfo {pages} {4716} (\bibinfo {year} {1994})}\BibitemShut {NoStop}%
\bibitem [{\citenamefont {Sorriso-Valvo}\ \emph {et~al.}(2002)\citenamefont
  {Sorriso-Valvo}, \citenamefont {Carbone}, \citenamefont {Noullez},
  \citenamefont {Politano}, \citenamefont {Pouquet},\ and\ \citenamefont
  {Veltri}}]{sorriso2002analysis}%
  \BibitemOpen
  \bibfield  {author} {\bibinfo {author} {\bibfnamefont {L.}~\bibnamefont
  {Sorriso-Valvo}}, \bibinfo {author} {\bibfnamefont {V.}~\bibnamefont
  {Carbone}}, \bibinfo {author} {\bibfnamefont {A.}~\bibnamefont {Noullez}},
  \bibinfo {author} {\bibfnamefont {H.}~\bibnamefont {Politano}}, \bibinfo
  {author} {\bibfnamefont {A.}~\bibnamefont {Pouquet}}, \ and\ \bibinfo
  {author} {\bibfnamefont {P.}~\bibnamefont {Veltri}},\ }\bibfield  {title}
  {\enquote {\bibinfo {title} {Analysis of cancellation in two-dimensional
  magnetohydrodynamic turbulence},}\ }\href@noop {} {\bibfield  {journal}
  {\bibinfo  {journal} {Phys. Plasmas}\ }\textbf {\bibinfo {volume} {9}},\
  \bibinfo {pages} {89--95} (\bibinfo {year} {2002})}\BibitemShut {NoStop}%
\bibitem [{\citenamefont {Graham}, \citenamefont {Mininni},\ and\ \citenamefont
  {Pouquet}(2005)}]{graham2005cancellation}%
  \BibitemOpen
  \bibfield  {author} {\bibinfo {author} {\bibfnamefont {J.~P.}\ \bibnamefont
  {Graham}}, \bibinfo {author} {\bibfnamefont {P.~D.}\ \bibnamefont {Mininni}},
  \ and\ \bibinfo {author} {\bibfnamefont {A.}~\bibnamefont {Pouquet}},\
  }\bibfield  {title} {\enquote {\bibinfo {title} {{Cancellation exponent and
  multifractal structure in two-dimensional magnetohydrodynamics: direct
  numerical simulations and Lagrangian averaged modeling}},}\ }\href@noop {}
  {\bibfield  {journal} {\bibinfo  {journal} {Phys. Rev. E}\ }\textbf {\bibinfo
  {volume} {72}},\ \bibinfo {pages} {045301} (\bibinfo {year}
  {2005})}\BibitemShut {NoStop}%
\bibitem [{\citenamefont {Martin}\ \emph {et~al.}(2013)\citenamefont {Martin},
  \citenamefont {De~Vita}, \citenamefont {Sorriso-Valvo}, \citenamefont
  {Dmitruk}, \citenamefont {Nigro}, \citenamefont {Primavera},\ and\
  \citenamefont {Carbone}}]{martin2013cancellation}%
  \BibitemOpen
  \bibfield  {author} {\bibinfo {author} {\bibfnamefont {L.}~\bibnamefont
  {Martin}}, \bibinfo {author} {\bibfnamefont {G.}~\bibnamefont {De~Vita}},
  \bibinfo {author} {\bibfnamefont {L.}~\bibnamefont {Sorriso-Valvo}}, \bibinfo
  {author} {\bibfnamefont {P.}~\bibnamefont {Dmitruk}}, \bibinfo {author}
  {\bibfnamefont {G.}~\bibnamefont {Nigro}}, \bibinfo {author} {\bibfnamefont
  {L.}~\bibnamefont {Primavera}}, \ and\ \bibinfo {author} {\bibfnamefont
  {V.}~\bibnamefont {Carbone}},\ }\bibfield  {title} {\enquote {\bibinfo
  {title} {Cancellation properties in hall magnetohydrodynamics with a strong
  guide magnetic field},}\ }\href@noop {} {\bibfield  {journal} {\bibinfo
  {journal} {Phys. Rev. E}\ }\textbf {\bibinfo {volume} {88}},\ \bibinfo
  {pages} {063107} (\bibinfo {year} {2013})}\BibitemShut {NoStop}%
\bibitem [{\citenamefont {Carbone}\ and\ \citenamefont
  {Bruno}(1997)}]{carbone1997sign}%
  \BibitemOpen
  \bibfield  {author} {\bibinfo {author} {\bibfnamefont {V.}~\bibnamefont
  {Carbone}}\ and\ \bibinfo {author} {\bibfnamefont {R.}~\bibnamefont
  {Bruno}},\ }\bibfield  {title} {\enquote {\bibinfo {title} {Sign singularity
  of the magnetic helicity from in situ solar wind observations},}\ }\href@noop
  {} {\bibfield  {journal} {\bibinfo  {journal} {Astrophys. J.}\ }\textbf
  {\bibinfo {volume} {488}},\ \bibinfo {pages} {482} (\bibinfo {year}
  {1997})}\BibitemShut {NoStop}%
\bibitem [{\citenamefont {Consolini}\ and\ \citenamefont
  {Lui}(1999)}]{consolini1999sign}%
  \BibitemOpen
  \bibfield  {author} {\bibinfo {author} {\bibfnamefont {G.}~\bibnamefont
  {Consolini}}\ and\ \bibinfo {author} {\bibfnamefont {A.~T.}\ \bibnamefont
  {Lui}},\ }\bibfield  {title} {\enquote {\bibinfo {title} {Sign-singularity
  analysis of current disruption},}\ }\href@noop {} {\bibfield  {journal}
  {\bibinfo  {journal} {{Geophys. Res. Lett.}}\ }\textbf {\bibinfo {volume}
  {26}},\ \bibinfo {pages} {1673--1676} (\bibinfo {year} {1999})}\BibitemShut
  {NoStop}%
\bibitem [{\citenamefont {Carbone}\ \emph {et~al.}(2010)\citenamefont
  {Carbone}, \citenamefont {Perri}, \citenamefont {Yordanova}, \citenamefont
  {Veltri}, \citenamefont {Bruno}, \citenamefont {Khotyaintsev},\ and\
  \citenamefont {Andr{\'e}}}]{carbone2010sign}%
  \BibitemOpen
  \bibfield  {author} {\bibinfo {author} {\bibfnamefont {V.}~\bibnamefont
  {Carbone}}, \bibinfo {author} {\bibfnamefont {S.}~\bibnamefont {Perri}},
  \bibinfo {author} {\bibfnamefont {E.}~\bibnamefont {Yordanova}}, \bibinfo
  {author} {\bibfnamefont {P.}~\bibnamefont {Veltri}}, \bibinfo {author}
  {\bibfnamefont {R.}~\bibnamefont {Bruno}}, \bibinfo {author} {\bibfnamefont
  {Y.}~\bibnamefont {Khotyaintsev}}, \ and\ \bibinfo {author} {\bibfnamefont
  {M.}~\bibnamefont {Andr{\'e}}},\ }\bibfield  {title} {\enquote {\bibinfo
  {title} {Sign-singularity of the reduced magnetic helicity in the solar wind
  plasma},}\ }\href@noop {} {\bibfield  {journal} {\bibinfo  {journal} {Phys.
  Rev. Lett.}\ }\textbf {\bibinfo {volume} {104}},\ \bibinfo {pages} {181101}
  (\bibinfo {year} {2010})}\BibitemShut {NoStop}%
\bibitem [{\citenamefont {Sorriso-Valvo}\ \emph {et~al.}(2015)\citenamefont
  {Sorriso-Valvo}, \citenamefont {De~Vita}, \citenamefont {Kazachenko},
  \citenamefont {Krucker}, \citenamefont {Primavera}, \citenamefont {Servidio},
  \citenamefont {Vecchio}, \citenamefont {Welsch}, \citenamefont {Fisher},
  \citenamefont {Lepreti},\ and\ \citenamefont {Carbone}}]{sorriso2015sign}%
  \BibitemOpen
  \bibfield  {author} {\bibinfo {author} {\bibfnamefont {L.}~\bibnamefont
  {Sorriso-Valvo}}, \bibinfo {author} {\bibfnamefont {G.}~\bibnamefont
  {De~Vita}}, \bibinfo {author} {\bibfnamefont {M.~D.}\ \bibnamefont
  {Kazachenko}}, \bibinfo {author} {\bibfnamefont {S.}~\bibnamefont {Krucker}},
  \bibinfo {author} {\bibfnamefont {L.}~\bibnamefont {Primavera}}, \bibinfo
  {author} {\bibfnamefont {S.}~\bibnamefont {Servidio}}, \bibinfo {author}
  {\bibfnamefont {A.}~\bibnamefont {Vecchio}}, \bibinfo {author} {\bibfnamefont
  {B.~T.}\ \bibnamefont {Welsch}}, \bibinfo {author} {\bibfnamefont {G.~H.}\
  \bibnamefont {Fisher}}, \bibinfo {author} {\bibfnamefont {F.}~\bibnamefont
  {Lepreti}}, \ and\ \bibinfo {author} {\bibfnamefont {V.}~\bibnamefont
  {Carbone}},\ }\bibfield  {title} {\enquote {\bibinfo {title} {Sign
  singularity and flares in solar active region noaa 11158},}\ }\href@noop {}
  {\bibfield  {journal} {\bibinfo  {journal} {Astrophys. J.}\ }\textbf
  {\bibinfo {volume} {801}},\ \bibinfo {pages} {36} (\bibinfo {year}
  {2015})}\BibitemShut {NoStop}%
\bibitem [{\citenamefont {De~Michelis}, \citenamefont {Consolini},\ and\
  \citenamefont {Meloni}(1998)}]{de1998sign}%
  \BibitemOpen
  \bibfield  {author} {\bibinfo {author} {\bibfnamefont {P.}~\bibnamefont
  {De~Michelis}}, \bibinfo {author} {\bibfnamefont {G.}~\bibnamefont
  {Consolini}}, \ and\ \bibinfo {author} {\bibfnamefont {A.}~\bibnamefont
  {Meloni}},\ }\bibfield  {title} {\enquote {\bibinfo {title} {Sign singularity
  in the secular acceleration of the geomagnetic field},}\ }\href@noop {}
  {\bibfield  {journal} {\bibinfo  {journal} {Phys. Rev. Lett.}\ }\textbf
  {\bibinfo {volume} {81}},\ \bibinfo {pages} {5023} (\bibinfo {year}
  {1998})}\BibitemShut {NoStop}%
\bibitem [{\citenamefont {Imazio}\ and\ \citenamefont
  {Mininni}(2010)}]{imazio2010cancellation}%
  \BibitemOpen
  \bibfield  {author} {\bibinfo {author} {\bibfnamefont {P.~R.}\ \bibnamefont
  {Imazio}}\ and\ \bibinfo {author} {\bibfnamefont {P.~D.}\ \bibnamefont
  {Mininni}},\ }\bibfield  {title} {\enquote {\bibinfo {title} {Cancellation
  exponents in helical and non-helical flows},}\ }\href@noop {} {\bibfield
  {journal} {\bibinfo  {journal} {J. Fluid Mech.}\ }\textbf {\bibinfo {volume}
  {651}},\ \bibinfo {pages} {241--250} (\bibinfo {year} {2010})}\BibitemShut
  {NoStop}%
\bibitem [{\citenamefont {Horne}\ and\ \citenamefont
  {Mininni}(2013)}]{horne2013sign}%
  \BibitemOpen
  \bibfield  {author} {\bibinfo {author} {\bibfnamefont {E.}~\bibnamefont
  {Horne}}\ and\ \bibinfo {author} {\bibfnamefont {P.~D.}\ \bibnamefont
  {Mininni}},\ }\bibfield  {title} {\enquote {\bibinfo {title} {Sign
  cancellation and scaling in the vertical component of velocity and vorticity
  in rotating turbulence},}\ }\href@noop {} {\bibfield  {journal} {\bibinfo
  {journal} {Phys. Rev. E}\ }\textbf {\bibinfo {volume} {88}},\ \bibinfo
  {pages} {013011} (\bibinfo {year} {2013})}\BibitemShut {NoStop}%
\bibitem [{\citenamefont {Mandelbrot}(1974)}]{mandelbrot}%
  \BibitemOpen
  \bibfield  {author} {\bibinfo {author} {\bibfnamefont {B.~B.}\ \bibnamefont
  {Mandelbrot}},\ }\bibfield  {title} {\enquote {\bibinfo {title} {Intermittent
  turbulence in self-similar cascades: divergence of high moments and dimension
  of the carrier},}\ }\href@noop {} {\bibfield  {journal} {\bibinfo  {journal}
  {J. Fluid Mech.}\ }\textbf {\bibinfo {volume} {62}},\ \bibinfo {pages}
  {331--358} (\bibinfo {year} {1974})}\BibitemShut {NoStop}%
\bibitem [{\citenamefont {Sreenivasan}(1991)}]{sreeni.1991}%
  \BibitemOpen
  \bibfield  {author} {\bibinfo {author} {\bibfnamefont {K.~R.}\ \bibnamefont
  {Sreenivasan}},\ }\bibfield  {title} {\enquote {\bibinfo {title} {Fractals
  and multifractals in fluid turbulence},}\ }\href@noop {} {\bibfield
  {journal} {\bibinfo  {journal} {Annu. Rev. Fluid Mech.}\ }\textbf {\bibinfo
  {volume} {23}},\ \bibinfo {pages} {539--600} (\bibinfo {year}
  {1991})}\BibitemShut {NoStop}%
\bibitem [{\citenamefont {Vainshtein}, \citenamefont {Du},\ and\ \citenamefont
  {Sreenivasan}(1994)}]{vainshtein1994sign}%
  \BibitemOpen
  \bibfield  {author} {\bibinfo {author} {\bibfnamefont {S.~I.}\ \bibnamefont
  {Vainshtein}}, \bibinfo {author} {\bibfnamefont {Y.}~\bibnamefont {Du}}, \
  and\ \bibinfo {author} {\bibfnamefont {K.~R.}\ \bibnamefont {Sreenivasan}},\
  }\bibfield  {title} {\enquote {\bibinfo {title} {Sign-singular measure and
  its association with turbulent scalings},}\ }\href@noop {} {\bibfield
  {journal} {\bibinfo  {journal} {Phys. Rev. E}\ }\textbf {\bibinfo {volume}
  {49}},\ \bibinfo {pages} {R2521} (\bibinfo {year} {1994})}\BibitemShut
  {NoStop}%
\bibitem [{\citenamefont {Sreenivasan}, \citenamefont {Juneja},\ and\
  \citenamefont {Suri}(1995)}]{sreenivasan1995scaling}%
  \BibitemOpen
  \bibfield  {author} {\bibinfo {author} {\bibfnamefont {K.~R.}\ \bibnamefont
  {Sreenivasan}}, \bibinfo {author} {\bibfnamefont {A.}~\bibnamefont {Juneja}},
  \ and\ \bibinfo {author} {\bibfnamefont {A.~K.}\ \bibnamefont {Suri}},\
  }\bibfield  {title} {\enquote {\bibinfo {title} {{Scaling properties of
  circulation in moderate-Reynolds-number turbulent wakes}},}\ }\href@noop {}
  {\bibfield  {journal} {\bibinfo  {journal} {Phys. Rev. Lett.}\ }\textbf
  {\bibinfo {volume} {75}},\ \bibinfo {pages} {433} (\bibinfo {year}
  {1995})}\BibitemShut {NoStop}%
\bibitem [{\citenamefont {Yeung}, \citenamefont {Zhai},\ and\ \citenamefont
  {Sreenivasan}(2015)}]{YZS.2015}%
  \BibitemOpen
  \bibfield  {author} {\bibinfo {author} {\bibfnamefont {P.~K.}\ \bibnamefont
  {Yeung}}, \bibinfo {author} {\bibfnamefont {X.~M.}\ \bibnamefont {Zhai}}, \
  and\ \bibinfo {author} {\bibfnamefont {K.~R.}\ \bibnamefont {Sreenivasan}},\
  }\bibfield  {title} {\enquote {\bibinfo {title} {Extreme events in
  computational turbulence},}\ }\href@noop {} {\bibfield  {journal} {\bibinfo
  {journal} {Proc. Nat. Acad. Sci.}\ }\textbf {\bibinfo {volume} {{112}(41)}},\
  \bibinfo {pages} {{12633--12638}} (\bibinfo {year} {2015})}\BibitemShut
  {NoStop}%
\bibitem [{\citenamefont {Nikora}, \citenamefont {Goring},\ and\ \citenamefont
  {Camussi}(2001)}]{nikora2001intermittency}%
  \BibitemOpen
  \bibfield  {author} {\bibinfo {author} {\bibfnamefont {V.}~\bibnamefont
  {Nikora}}, \bibinfo {author} {\bibfnamefont {D.}~\bibnamefont {Goring}}, \
  and\ \bibinfo {author} {\bibfnamefont {R.}~\bibnamefont {Camussi}},\
  }\bibfield  {title} {\enquote {\bibinfo {title} {Intermittency and
  interrelationships between turbulence scaling exponents: Phase-randomization
  tests},}\ }\href@noop {} {\bibfield  {journal} {\bibinfo  {journal} {Phys.
  Fluids}\ }\textbf {\bibinfo {volume} {13}},\ \bibinfo {pages} {1404--1414}
  (\bibinfo {year} {2001})}\BibitemShut {NoStop}%
\bibitem [{\citenamefont {Zhai}\ and\ \citenamefont {Yeung}(2018)}]{ZY.2018}%
  \BibitemOpen
  \bibfield  {author} {\bibinfo {author} {\bibfnamefont {X.~M.}\ \bibnamefont
  {Zhai}}\ and\ \bibinfo {author} {\bibfnamefont {P.~K.}\ \bibnamefont
  {Yeung}},\ }\bibfield  {title} {\enquote {\bibinfo {title} {The evolution of
  anisotropy in direct numerical simulations of mhd turbulence in a strong
  magnetic field on elongated periodic domains},}\ }\href@noop {} {\bibfield
  {journal} {\bibinfo  {journal} {Phys. Rev. Fluids.}\ }\textbf {\bibinfo
  {volume} {3}},\ \bibinfo {pages} {084602} (\bibinfo {year}
  {2018})}\BibitemShut {NoStop}%
\bibitem [{\citenamefont {Jimenez}\ \emph {et~al.}(1993)\citenamefont
  {Jimenez}, \citenamefont {Wray}, \citenamefont {Saffman},\ and\ \citenamefont
  {Rogallo}}]{JWSR1993}%
  \BibitemOpen
  \bibfield  {author} {\bibinfo {author} {\bibfnamefont {J.}~\bibnamefont
  {Jimenez}}, \bibinfo {author} {\bibfnamefont {A.~A.}\ \bibnamefont {Wray}},
  \bibinfo {author} {\bibfnamefont {P.~G.}\ \bibnamefont {Saffman}}, \ and\
  \bibinfo {author} {\bibfnamefont {R.~S.}\ \bibnamefont {Rogallo}},\
  }\bibfield  {title} {\enquote {\bibinfo {title} {The structure of intense
  vorticity in isotropic turbulence},}\ }\href@noop {} {\bibfield  {journal}
  {\bibinfo  {journal} {J. Fluid Mech.}\ }\textbf {\bibinfo {volume} {{255}}},\
  \bibinfo {pages} {{65--90}} (\bibinfo {year} {1993})}\BibitemShut {NoStop}%
\bibitem [{\citenamefont {Davidson}, \citenamefont {Kaneda},\ and\
  \citenamefont {Sreenivasan}(2012)}]{tenchapters}%
  \BibitemOpen
  \bibfield  {author} {\bibinfo {author} {\bibfnamefont {P.~A.}\ \bibnamefont
  {Davidson}}, \bibinfo {author} {\bibfnamefont {Y.}~\bibnamefont {Kaneda}}, \
  and\ \bibinfo {author} {\bibfnamefont {K.~R.}\ \bibnamefont {Sreenivasan}},\
  }\href@noop {} {\emph {\bibinfo {title} {Ten chapters in turbulence}}}\
  (\bibinfo  {publisher} {Cambridge University Press},\ \bibinfo {year}
  {2012})\BibitemShut {NoStop}%
\bibitem [{\citenamefont {Eswaran}\ and\ \citenamefont {Pope}(1988)}]{EP88}%
  \BibitemOpen
  \bibfield  {author} {\bibinfo {author} {\bibfnamefont {V.}~\bibnamefont
  {Eswaran}}\ and\ \bibinfo {author} {\bibfnamefont {S.~B.}\ \bibnamefont
  {Pope}},\ }\bibfield  {title} {\enquote {\bibinfo {title} {An examination of
  forcing in direct numerical simulations of turbulence},}\ }\href@noop {}
  {\bibfield  {journal} {\bibinfo  {journal} {Comput. Fluids}\ }\textbf
  {\bibinfo {volume} {16}},\ \bibinfo {pages} {257--278} (\bibinfo {year}
  {1988})}\BibitemShut {NoStop}%
\bibitem [{\citenamefont {Donzis}\ and\ \citenamefont {Yeung}(2010)}]{DY2010}%
  \BibitemOpen
  \bibfield  {author} {\bibinfo {author} {\bibfnamefont {D.~A.}\ \bibnamefont
  {Donzis}}\ and\ \bibinfo {author} {\bibfnamefont {P.~K.}\ \bibnamefont
  {Yeung}},\ }\bibfield  {title} {\enquote {\bibinfo {title} {Resolution
  effects and scaling in numerical simulations of passive scalar mixing in
  turbulence},}\ }\href@noop {} {\bibfield  {journal} {\bibinfo  {journal}
  {Physica D}\ }\textbf {\bibinfo {volume} {{239}}},\ \bibinfo {pages}
  {{1278--1287}} (\bibinfo {year} {2010})}\BibitemShut {NoStop}%
\bibitem [{\citenamefont {Rogallo}(1981)}]{rogallo}%
  \BibitemOpen
  \bibfield  {author} {\bibinfo {author} {\bibfnamefont {R.~S.}\ \bibnamefont
  {Rogallo}},\ }\bibfield  {title} {\enquote {\bibinfo {title} {Numerical
  experiments in homogeneous turbulence. {{\em NASA Tech. Memo. 81315}, NASA
  Ames Research Center}},}\ }\href@noop {} {\  (\bibinfo {year}
  {1981})}\BibitemShut {NoStop}%
\bibitem [{\citenamefont {Ishihara}, \citenamefont {Gotoh},\ and\ \citenamefont
  {Kaneda}(2009)}]{IGK2009}%
  \BibitemOpen
  \bibfield  {author} {\bibinfo {author} {\bibfnamefont {T.}~\bibnamefont
  {Ishihara}}, \bibinfo {author} {\bibfnamefont {T.}~\bibnamefont {Gotoh}}, \
  and\ \bibinfo {author} {\bibfnamefont {Y.}~\bibnamefont {Kaneda}},\
  }\bibfield  {title} {\enquote {\bibinfo {title} {Study of high-{R}eynolds
  number isotropic turbulence by direct numerical simulation},}\ }\href@noop {}
  {\bibfield  {journal} {\bibinfo  {journal} {Annu. Rev. Fluid Mech.}\ }\textbf
  {\bibinfo {volume} {41}},\ \bibinfo {pages} {165--180} (\bibinfo {year}
  {2009})}\BibitemShut {NoStop}%
\bibitem [{\citenamefont {Yeung}, \citenamefont {Sreenivasan},\ and\
  \citenamefont {Pope}(2018)}]{YSP.2018}%
  \BibitemOpen
  \bibfield  {author} {\bibinfo {author} {\bibfnamefont {P.~K.}\ \bibnamefont
  {Yeung}}, \bibinfo {author} {\bibfnamefont {K.~R.}\ \bibnamefont
  {Sreenivasan}}, \ and\ \bibinfo {author} {\bibfnamefont {S.~B.}\ \bibnamefont
  {Pope}},\ }\bibfield  {title} {\enquote {\bibinfo {title} {Effects of finite
  spatial and temporal resolution in direct numerical simulations of
  incompressible isotropic turbulence},}\ }\href@noop {} {\bibfield  {journal}
  {\bibinfo  {journal} {Phys. Rev. Fluids.}\ }\textbf {\bibinfo {volume} {3}},\
  \bibinfo {pages} {064603} (\bibinfo {year} {2018})}\BibitemShut {NoStop}%
\bibitem [{\citenamefont {Donzis}, \citenamefont {Yeung},\ and\ \citenamefont
  {Pekurovsky}(2008)}]{DYP2008}%
  \BibitemOpen
  \bibfield  {author} {\bibinfo {author} {\bibfnamefont {D.~A.}\ \bibnamefont
  {Donzis}}, \bibinfo {author} {\bibfnamefont {P.~K.}\ \bibnamefont {Yeung}}, \
  and\ \bibinfo {author} {\bibfnamefont {D.}~\bibnamefont {Pekurovsky}},\
  }\bibfield  {title} {\enquote {\bibinfo {title} {Turbulence simulations on
  ${O}(10^4)$ processors},}\ }in\ \href@noop {} {\emph {\bibinfo {booktitle}
  {Proc. {TeraGrid} '08 Conf {\rm (Las Vegas, NV)}}}}\ (\bibinfo {year}
  {2008})\BibitemShut {NoStop}%
\bibitem [{\citenamefont {Iyer}(2014)}]{iyer.thesis}%
  \BibitemOpen
  \bibfield  {author} {\bibinfo {author} {\bibfnamefont {K.~P.}\ \bibnamefont
  {Iyer}},\ }\emph {\bibinfo {title} {Studies of turbulence structure and
  turbulent mixing using Petascale computing}},\ \href@noop {} {Ph.D. thesis},\
  \bibinfo  {school} {Georgia Institute of Technology} (\bibinfo {year}
  {2014})\BibitemShut {NoStop}%
\bibitem [{\citenamefont {Hentschel}\ and\ \citenamefont
  {Procaccia}(1983)}]{hentschel1983}%
  \BibitemOpen
  \bibfield  {author} {\bibinfo {author} {\bibfnamefont {H.}~\bibnamefont
  {Hentschel}}\ and\ \bibinfo {author} {\bibfnamefont {I.}~\bibnamefont
  {Procaccia}},\ }\bibfield  {title} {\enquote {\bibinfo {title} {The infinite
  number of generalized dimensions of fractals and strange attractors},}\
  }\href@noop {} {\bibfield  {journal} {\bibinfo  {journal} {Physica D}\
  }\textbf {\bibinfo {volume} {8}},\ \bibinfo {pages} {435--444} (\bibinfo
  {year} {1983})}\BibitemShut {NoStop}%
\bibitem [{\citenamefont {Iyer}, \citenamefont {Sreenivasan},\ and\
  \citenamefont {Yeung}(2017)}]{ISY.2017}%
  \BibitemOpen
  \bibfield  {author} {\bibinfo {author} {\bibfnamefont {K.~P.}\ \bibnamefont
  {Iyer}}, \bibinfo {author} {\bibfnamefont {K.~R.}\ \bibnamefont
  {Sreenivasan}}, \ and\ \bibinfo {author} {\bibfnamefont {P.~K.}\ \bibnamefont
  {Yeung}},\ }\bibfield  {title} {\enquote {\bibinfo {title} {Reynolds number
  scaling of velocity increments in isotropic turbulence},}\ }\href@noop {}
  {\bibfield  {journal} {\bibinfo  {journal} {Phys. Rev. E}\ }\textbf {\bibinfo
  {volume} {95}},\ \bibinfo {pages} {021101} (\bibinfo {year}
  {2017})}\BibitemShut {NoStop}%
\bibitem [{\citenamefont {Kolmogorov}(1991)}]{K41}%
  \BibitemOpen
  \bibfield  {author} {\bibinfo {author} {\bibfnamefont {A.~N.}\ \bibnamefont
  {Kolmogorov}},\ }\bibfield  {title} {\enquote {\bibinfo {title} {{The local
  structure of turbulence in an incompressible fluid with very large Reynolds
  numbers}},}\ }\href@noop {} {\bibfield  {journal} {\bibinfo  {journal} {{\em
  Dokl. Akad. Nauk SSSR}}\ }\textbf {\bibinfo {volume} {30}},\ \bibinfo {pages}
  {301--305} (\bibinfo {year} {1941, Reprinted as {\em Proc. Roy Soc. Lond.
  A.}, 434:9-13, 1991})}\BibitemShut {NoStop}%
\bibitem [{\citenamefont {Kolmogorov}(1962)}]{K62}%
  \BibitemOpen
  \bibfield  {author} {\bibinfo {author} {\bibfnamefont {A.~N.}\ \bibnamefont
  {Kolmogorov}},\ }\bibfield  {title} {\enquote {\bibinfo {title} {{A
  refinement of previous hypotheses concerning the local structure of
  turbulence in a viscous incompressible fluid at high Reynolds number}},}\
  }\href@noop {} {\bibfield  {journal} {\bibinfo  {journal} {J. Fluid Mech.}\
  }\textbf {\bibinfo {volume} {{13}}},\ \bibinfo {pages} {{82--85}} (\bibinfo
  {year} {1962})}\BibitemShut {NoStop}%
\bibitem [{\citenamefont {Sreenivasan}\ and\ \citenamefont
  {Kailasnatth}(1993)}]{SK.1993}%
  \BibitemOpen
  \bibfield  {author} {\bibinfo {author} {\bibfnamefont {K.~R.}\ \bibnamefont
  {Sreenivasan}}\ and\ \bibinfo {author} {\bibfnamefont {P.}~\bibnamefont
  {Kailasnatth}},\ }\bibfield  {title} {\enquote {\bibinfo {title} {An update
  on the intermittency exponent in turbulence},}\ }\href@noop {} {\bibfield
  {journal} {\bibinfo  {journal} {Phys. Fluids A}\ }\textbf {\bibinfo {volume}
  {5}},\ \bibinfo {pages} {{2766--2769}} (\bibinfo {year} {1993})}\BibitemShut
  {NoStop}%
\bibitem [{\citenamefont {Donzis}\ and\ \citenamefont
  {Sreenivasan}(2010)}]{DS2010b}%
  \BibitemOpen
  \bibfield  {author} {\bibinfo {author} {\bibfnamefont {D.~A.}\ \bibnamefont
  {Donzis}}\ and\ \bibinfo {author} {\bibfnamefont {K.~R.}\ \bibnamefont
  {Sreenivasan}},\ }\bibfield  {title} {\enquote {\bibinfo {title} {The
  bottleneck effect and the kolmogorov constant in isotropic turbulence},}\
  }\href@noop {} {\bibfield  {journal} {\bibinfo  {journal} {J. Fluid Mech.}\
  }\textbf {\bibinfo {volume} {657}},\ \bibinfo {pages} {171--188} (\bibinfo
  {year} {2010})}\BibitemShut {NoStop}%
\bibitem [{\citenamefont {Zikanov}\ and\ \citenamefont
  {Thess}(1998)}]{zikanov.1998}%
  \BibitemOpen
  \bibfield  {author} {\bibinfo {author} {\bibfnamefont {O.}~\bibnamefont
  {Zikanov}}\ and\ \bibinfo {author} {\bibfnamefont {A.}~\bibnamefont
  {Thess}},\ }\bibfield  {title} {\enquote {\bibinfo {title} {Direct numerical
  simulation of forced {MHD} turbulence at low magnetic {Reynolds} number},}\
  }\href@noop {} {\bibfield  {journal} {\bibinfo  {journal} {J. Fluid Mech.}\
  }\textbf {\bibinfo {volume} {358}},\ \bibinfo {pages} {299--333} (\bibinfo
  {year} {1998})}\BibitemShut {NoStop}%
\bibitem [{\citenamefont {Reddy}\ and\ \citenamefont
  {Verma}(2014)}]{reddy.2014b}%
  \BibitemOpen
  \bibfield  {author} {\bibinfo {author} {\bibfnamefont {K.~S.}\ \bibnamefont
  {Reddy}}\ and\ \bibinfo {author} {\bibfnamefont {M.~K.}\ \bibnamefont
  {Verma}},\ }\bibfield  {title} {\enquote {\bibinfo {title} {Strong anisotropy
  in quasi-static magnetohydrodynamic turbulence for high interaction
  parameters},}\ }\href@noop {} {\bibfield  {journal} {\bibinfo  {journal}
  {Phys. Fluids}\ }\textbf {\bibinfo {volume} {26}},\ \bibinfo {pages} {025109}
  (\bibinfo {year} {2014})}\BibitemShut {NoStop}%
\end{thebibliography}

%

\end{document}